\documentclass[letter]{aa}
\usepackage{graphicx}
\usepackage{natbib,txfonts}
\usepackage[colorlinks=true, citecolor=blue]{hyperref}

\bibpunct{(}{)}{;}{a}{}{,} 

\def\kms{km\,s$^{-1}$}
\def\teff{$\mathrm{\textit{T}_{\text{eff}}}$}
\def\logg{\text{log}(\textit{g})}
\def\mh{[\text{M}/\text{H}]}
\def\feh{[\text{Fe}/\text{H}]}
\def\alpham{[$\alpha$/\text{M}]}
\def\alphafe{[$\alpha$/\text{Fe}]}
\def\rg{$\mathrm{R_{g}}$}
\def\snr{S/N}

\def\g{G}
\def\bp{$\mathrm{\text{B}_\text{p}}$}
\def\rp{$\mathrm{\text{R}_\text{p}}$}

\title{Insights from Super-Metal-Rich Stars:\\ Is the Milky Way bar young?}
\titlerunning{Timing Milky Way bar formation}

\authorrunning{Nepal et al.}
\author{S.~Nepal \inst{1, 2},
C.~Chiappini \inst{1},
G. Guiglion \inst{3,4},
M. Steinmetz \inst{1},
A. P\'erez-Villegas \inst{5}, \\
A. B. Queiroz \inst{1},
A. Miglio \inst{6},
P. Dohme \inst{7},
A. Khalatyan \inst{1}}
\institute{{Leibniz-Institut f\"ur Astrophysik Potsdam (AIP), An der Sternwarte 
16, 14482 Potsdam, Germany} \\
\email{snepal@aip.de}
\and
{Institut f\"ur Physik und Astronomie, Universit\"at Potsdam, Karl-Liebknecht-Str. 24/25, 14476 Potsdam, Germany}
\and
{Zentrum f\"ur Astronomie der Universit\"at Heidelberg, Landessternwarte, K\"onigstuhl 12, 69117 Heidelberg, Germany}
\and
{Max Planck Institute for Astronomy, K\"onigstuhl 17, 69117, Heidelberg, Germany}
\and
{Instituto de Astronom\'ia, Universidad Nacional Aut\'onoma de M\'exico, A. P. 106, C.P. 22800 Ensenada, B. C., M\'exico}
\and
{Dipartimento di Fisica e Astronomia, Universit\`a degli Studi di Bologna, Via Gobetti 93/2, I-40129 Bologna, Italy}
\and
{Freie Universit\"at Berlin, Fachbereich Physik, Arnimallee 14, 14195 Berlin, Germany}
}

\date{Received 23 October 2023 / accepted 28 November 2023}

\begin{document}

\abstract{Bar formation and merger events can contribute to the rearrangement of stars within the Galaxy in addition to triggering star formation (SF) epochs. Super-metal-rich (SMR) stars found at the Solar Neighbourhood (SNd) can be used as tracers of such events as they are expected to originate only in the inner Galaxy and have definitely migrated.}
{We aim at studying a homogeneous and large sample of SMR stars in the SNd to provide tighter constraints on the epoch of the bar formation and its impact on the Milky Way (MW) disc stellar populations.}
{We investigate a sample of 169\,701 main sequence turnoff (MSTO) and sub-giant branch (SGB) stars with 6D phase space information and high-quality stellar parameters coming from the {\tt hybrid-CNN} analysis of the \emph{Gaia}-DR3 RVS stars. We compute distances and ages using the {\tt StarHorse} code with a mean precision of 1\% and 11\%, respectively. From these, 11\,848 stars have metallicity (\feh) above 0.15 dex.}
{We report a metallicity dependence of spatial distribution of stellar orbits shown by the bimodal distribution in the guiding radius (\rg) at 6.9 and 7.9 kpc, first appearing at \feh$\sim$0.1 dex, becoming very pronounced at larger \feh. In addition, we've observed a trend where the most metal-rich stars, with \feh$\sim$0.4 dex, are predominantly old (9-12 Gyrs) but show a gradual decline in \feh\, with age, reaching around 0.25 dex at about 4 Gyrs ago, followed by a sharp drop around 3 Gyrs ago. Furthermore, our full dataset reveals a clear peak in the age-metallicity relationship during the same period, indicating a SF burst around 3-4 Gyrs ago with slightly sub-solar \feh\, and enhanced \alphafe.}
{We show the SMR stars are good tracers of the bar activity. We interpret the steep decrease in number of SMR stars at around 3 Gyr as the end of the bar formation epoch. In this scenario, the peak of bar activity also coincides with a peak in the SF activity in the disc. Although the SF burst around 3 Gyr ago has been reported previously, its origin was unclear. Here, we suggest the SF burst to have been triggered by the high bar activity, 3-4 Gyr ago. According to these results and interpretation, the MW bar could be young.}


\maketitle

\section{Introduction}

Stars are the luminous storytellers of our Galactic saga. By studying the properties of these stars, their ages, chemical abundances and motions, we can trace back the history of our Galaxy \citep{Pagel1997, Matteucci2001, Matteucci2021, freeman2002}. Among the different stellar populations in the Milky Way (MW), super-metal-rich (SMR\footnote{SMR stars are defined such that their metal-abundance exceeds the metallicity of the local present-day interstellar medium, and because of the negative radial abundance gradient in the MW, its value varies with galactocentric distance.}) stars are very interesting as they are expected to be formed only in the inner regions of our Galaxy out of materials enriched by previous generations of stars \citep{Grenon1972, Steinmetz1994, trevisan_2011, casagrande_2011, kordopatis2015, anders_2017, Miglio2021}. These SMR stars currently residing in the SNd have definitely migrated from the inner Galaxy \citep{Feuillet2018, Chen2019, Dantas2023} and their study can inform us about the processes that bring them to the outer disc. Indeed, \cite{Queiroz2021} have shown the inner kpcs of the Galaxy to provide a large reservoir of high-metallicity stars (either as part of an inner-thin disc or on bar shape orbits), and these are primary good candidates for migration.

The Galactic bar is considered an important perturber for generating radial motion of stars and gas in the galactic disc \cite[e.g.][]{Sellwood2014RvMP}. According to N-body simulations, during the formation and the phase of strong bar activity, stars in the disc are significantly redistributed, with the highest probability of migration at the bar resonances \citep{Minchev2010, Halle2015, Khoperskov2020_bar_escapees}. However, an open question still remains on the epoch of MW bar formation.

Old to young formation times for the MW bar have been proposed in both observational and simulation studies conducted so far. \citet{Bovy2019} and \citet{Wylie2022}, using red giant stars in the inner Galaxy with ages estimated using {\tt astroNN} \citep{leung2019b}, proposed an older bar at least 7 to 8 Gyr old based on the mean and peak of the age distribution. \citet{Cole2002}, using Carbon stars that trace the bar, suggested that bar likely formed $\sim$3 Gyr ago. However, it is important to recognise that the bar is a cumulative entity, it will indiscriminately contain older stars and stars born during bar formation. So, care has to be taken to differentiate older stars from those formed during bar formation (e.g. see \citealt{SaFreitas2023}). \citet{TepperGarcia2021arXiv}, using a tailored N-body model of barred MW, suggested that the bar formed 3-4 Gyrs ago. In context of external galaxies, \citet{Camila2023} recently reported the discovery of young bars, formed 4.5 and 0.7 Gyr ago, for spiral galaxies NGC 289 and NGC 1566 using the SF history of nuclear discs. This suggests that some disc galaxies, with stellar mass comparable to MW, could settle on longer time-scales.

However, because SMR stars are rare, we still lack a statistically significant sample to constrain this process. This situation has radically changed thanks to third data release (DR3) of the ESA \emph{Gaia} mission \citep{gaiadr3_survey_properties}. \emph{Gaia}-DR3 has provided about one million spectra from the Radial Velocity Spectrometer (RVS)\footnote{The RVS spectra were originally analysed during \emph{Gaia} DR3 (10.17876/gaia/dr.3) by the General Stellar Parametriser for spectroscopy (GSP-Spec, \citealt{recioblanco2023}) module of the Astrophysical parameters inference system (Apsis, \citealt{Creevey2023}).} of which only $\sim$178,000 stars have the good quality stellar parameters after the applying the recommended flags \citep{recioblanco2023}. A large portion of the published RVS spectra are of low signal-to-noise (\snr; 15 - 25 per pixel) and can be challenging for traditional spectroscopic pipelines. \citet[][G23]{rvs_cnn_2023} using a hybrid Convolutional Neural-Network (CNN) method, reanalyzed the RVS sample to derive atmospheric
parameters (\teff, \logg, and overall \mh) and chemical abundances
(\feh\, and \alpham), by supplementing extra information from \emph{Gaia} magnitudes \citep{Riello2021}, parallaxes \citep{Lindegren2021} and XP coefficients \citep{denageli_2023}. The {\tt hybrid-CNN} was trained with a high-quality training sample based on APOGEE DR17 \citep{Abdurrouf2022} labels and show precision and accuracy comparable to external data-sets such as GALAH and asteroseismology. Thanks to the novel method, G23 significantly improved the number of reliable targets that can be used for Galactic archaeology. The G23 catalog, has a large number of SMR stars, including a sample of MSTO and SGB stars, enabling the use of these traces to constrain the epoch of MW bar formation.

In this Letter, we explore the age-\feh\,(AMR) and the age-\alphafe\,(AAR) relationships along with the chemo-dynamics for a sample of MSTO and SGB stars selected from the RVS-CNN catalog of G23. In Sect. \ref{data}, we describe our sample focusing on the methods used to obtain the stellar ages and kinematics. In Sect. \ref{Section: resutls} we present our results and in Sect. \ref{Section:conclusion} we present the main conclusions.

\section{Data} \label{data}

We obtain a sample of stars, with $-$0.7 < \feh < 0.5 including a large set of the SMR stars, from the catalog of G23. We selected MSTO and SGB stars (see \citealt{Queiroz2023}) after the application of the G23 recommended flags. We keep only the MSTO-SGB stars as the stellar ages from isochrone fitting methods are most reliable for these evolutionary stages \cite[e.g.][]{Soderblom2010ARA}. We computed the distances and stellar ages with the {\tt StarHorse} Bayesian isochrone-fitting method \citep{queiroz2018, Queiroz2023} and integrate the orbits of the stars using {\tt Galpy} \citep{galpy2015}. For details on these computations see Appendix \ref{SH and galpy}.

We selected stars with relative age uncertainty less than 20\%, distance uncertainty less than 5\% and extinction uncertainty less than 0.2 Mag. We are left with a sample with mean uncertainty of 11\% for age and 1\% for distance due to very low parallax errors in the extended SNd thanks to \emph{Gaia}. We further remove any stars with poor astrometric solutions by limiting RUWE < 1.4 and also remove known variable stars by using \emph{Gaia} flag 'phot\_variable\_flag'$\neq$'VARIABLE' (see \citealt{gaiadr3_survey_properties}).

\begin{figure}[!ht]
    \centering
    \includegraphics[width=0.99\linewidth]{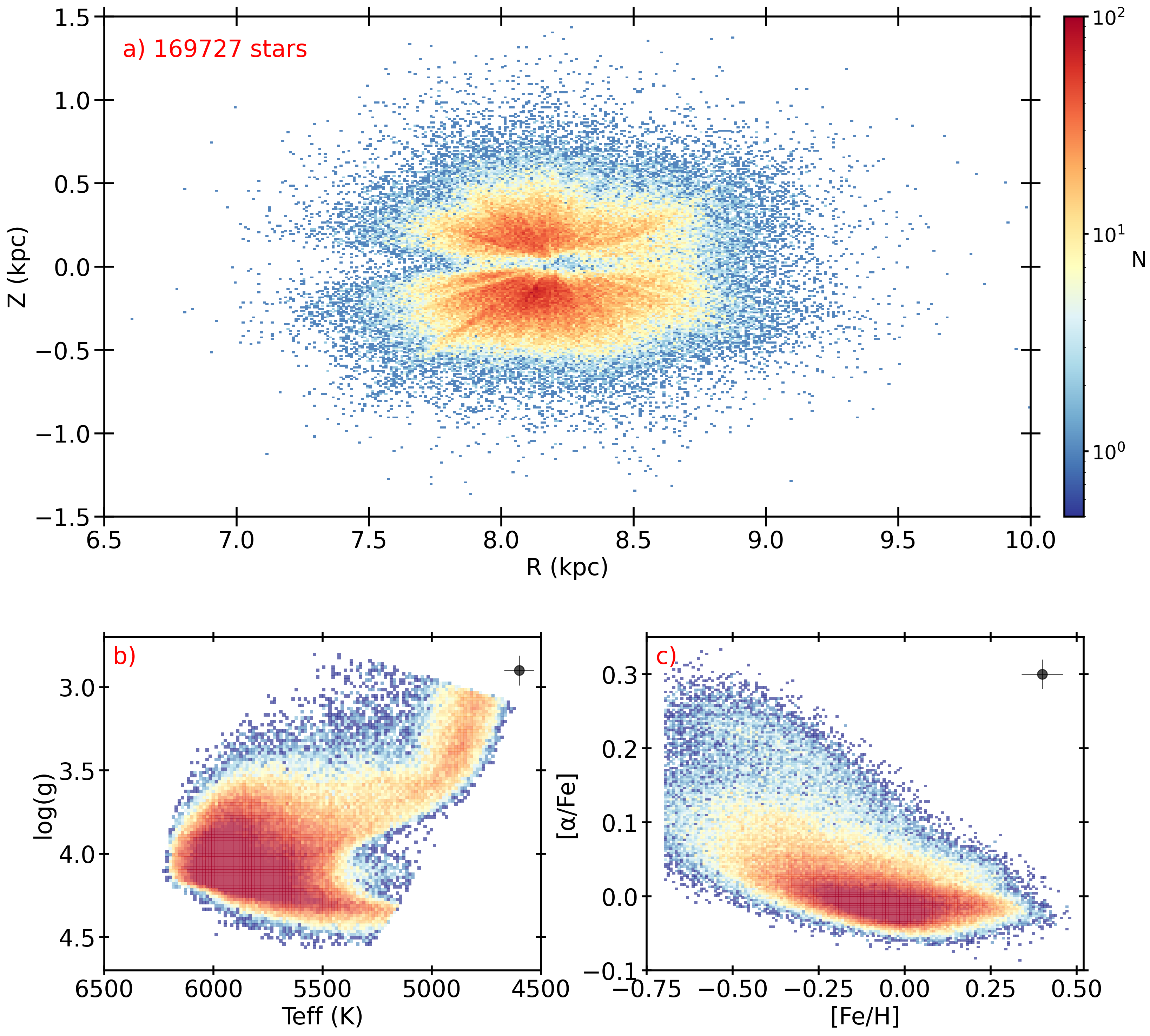}
    \caption{The properties of our selected sample - a) Distance from the galactic mid-plane (Z) vs the Galactocentric distance (R); b) the Kiel diagram (\logg\, vs \teff); and c) the \alphafe\, vs \feh\, diagram for the sample, colored by number density.}
    \label{fig:the_sample}
\end{figure}

This gives us a final sample of 169\,727 stars, including 19\,367 stars with \feh>0.1 dex. As a test, we apply the 13 recommended flags on the GSP-Spec parameters \citep{recioblanco2023} to find a total sample reduced to 20\,269 stars with only 4\,853 stars with \feh>0.1 dex.

In Fig.\ref{fig:the_sample}, we present our sample properties. Although the stars are widely distributed in Z vs R space, the number of stars decreases as we move away from SNd, as expected for a sample of MSTO+SGB stars (see \citealt{Queiroz2023}). Therefore the sample is essentially dominated by thin disc stars (low-\alphafe\, population). 

\begin{figure*}[!ht]
    \centering
    \includegraphics[width=0.8\linewidth]{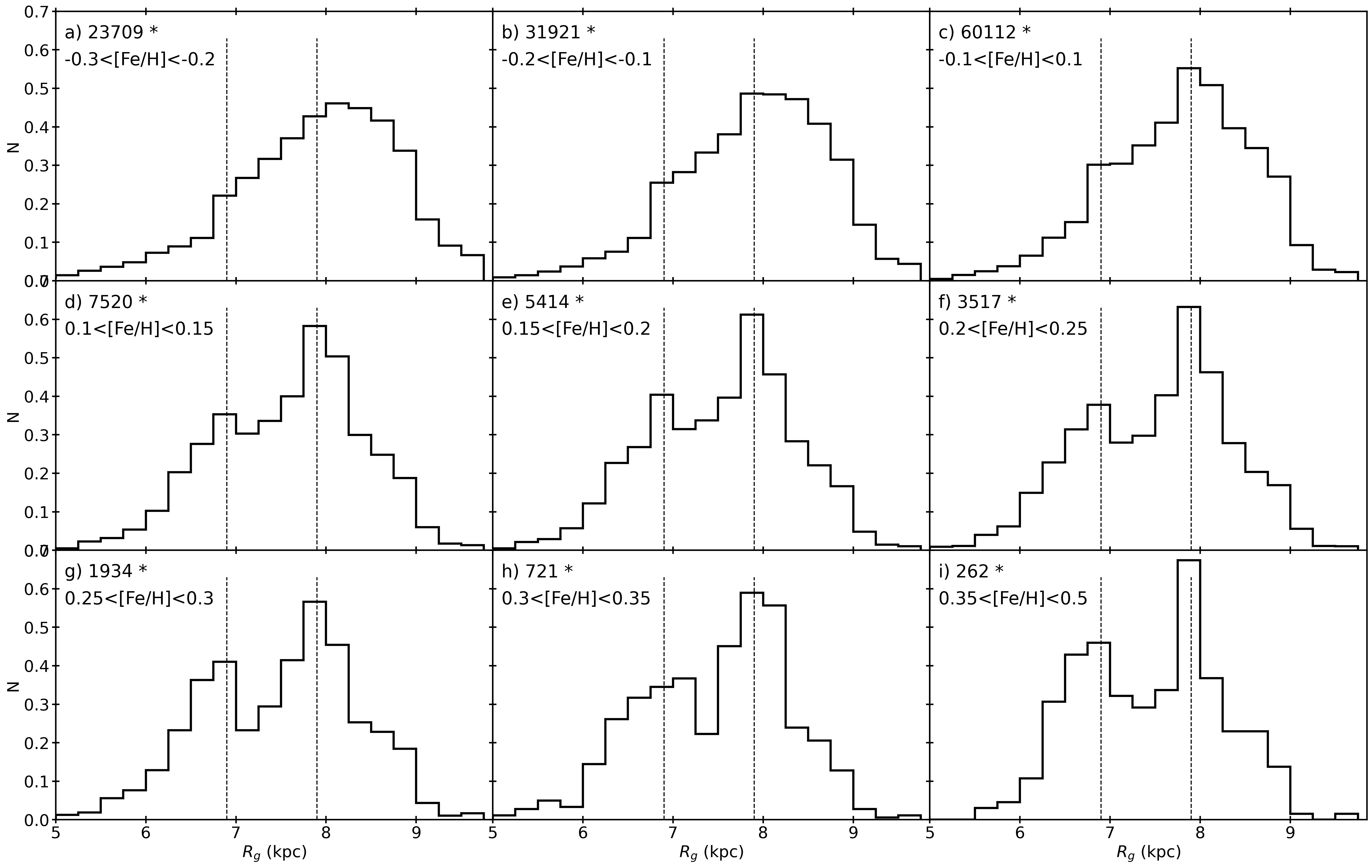}
    \caption{Distribution of the guiding radii (\rg) in bins of metallicity (\feh). The \feh\, values increase from top left to bottom right, i.e. from panel a) to panel i) (number of stars and the \feh\, range shown). The two dotted lines at 6.9 and 7.9 kpc represent the two peaks of the \rg\, distribution.}
    \label{fig:rguiding_bimodality}
\end{figure*}

\section{Results and Discussion} \label{Section: resutls}
 
\subsection{Super-metal-rich stars as tracers of MW bar activity}\label{bimodality}

Figure \ref{fig:rguiding_bimodality} shows the distribution of guiding radius (\rg), in bins of metallicity, for stars in the range $-$0.3<\feh<0.5. The \feh\, values increase as we move from panel a) to panel i). These histograms show that a clear bimodality in the distribution of \rg\, appears as we move from sub-solar \feh\, to SMR stars. The bimodality visibly appear first in the bin 0.10<\feh<0.15, as seen in panel (d), and becomes stronger for the higher \feh\, bins. An inner and an outer peak is seen at 6.9 and 7.9 kpc, for all super-solar bins. We validate this finding with an external sample of red-giant stars in Appendix \ref{seismic ages}. We note that the two peaks are not equally populated, with higher number of stars near the solar radius, and attribute this to the \emph{Gaia} selection function.
 
Unlike SMR stars, sub-solar to solar \feh\, stars are equally plausible to be formed in the SNd as well as the inner and outer galaxy, so their actual birth location can be uncertain due to radial migration (e.g. \citealt{minchev_2013, minchev_2014} and references therein). We also know the innermost regions of the MW to be populated of very metal-rich stars (e.g see \citealt{Barbuy2018ARA} and references therein). In particular, \cite{Queiroz2021}, using APOGEE and \emph{Gaia} surveys to study bar and bulge region, found a large cache of metal-rich stars in inner Galaxy. These metal-rich stars, found in cold to highly eccentric orbits, could be a source of origin for our local SMR stars.

What could be the mechanism that brought them here? The SMR stars in our sample, most of them in cold orbits with 96\% with ecc<0.3, distributed favourably at certain \rg\, entices us to view their distribution in context of evolution and dynamical process in the Galaxy (See Appendix \ref{kine} for a comparison of velocity dispersion of metal-rich stars with the full sample.). The Galactic bar has been long considered an efficient source of radial migration \citep{Combes1981, Pfenniger1990}. Galaxy models and N-body simulations have shown that slow down of Galactic bars can lead to migration of stars from inner to outer galaxy by trapping them in resonances as they travel outwards in the disc \citep[e.g.][]{Athanassoula2003, Khoperskov2020_bar_escapees, Chiba2021}.

Although dependent on the exact dynamical recipe, many recent studies have placed bar Corotation (CR) between radius of 6-7 kpc (e.g. \citealt{Portail2017, Khoperskov2020_bar_arms, Chiba2021}) and the Local arm at around 8 kpc (e.g. \citealt{Palicio2023}) linking it to Outer Lindbald Resonances \citep[OLR;][]{Fragkoudi2019MNRAS, Khoperskov2020_bar_arms}. Furthermore, \citet{Chen2022} using SMR stars, find ridges and undulations in the $\phi$ vs $L_Z$ plane similar to the orbits trapped in resonances of a slow bar as in the model of \cite{Monari2019}. See Appendix \ref{bar tracers} for test on how metal-rich stars are better suitable for probing bar activity. Therefore, the inner and outer \rg\, peaks, observed for the SMR stars at 6.9 and 7.9 kpc, most probably correspond to the imprints of the Galactic bar.

From our observations (see Fig.\ref{fig:rguiding_bimodality}) we conclude that the SMR stars, currently in the SNd, trace the signatures of bar activity. In the next subsection we employ these SMR stars, with good ages, to study the formation epoch of the MW bar.

\begin{figure*}[!ht]
    \centering
    \includegraphics[width=0.85\linewidth]{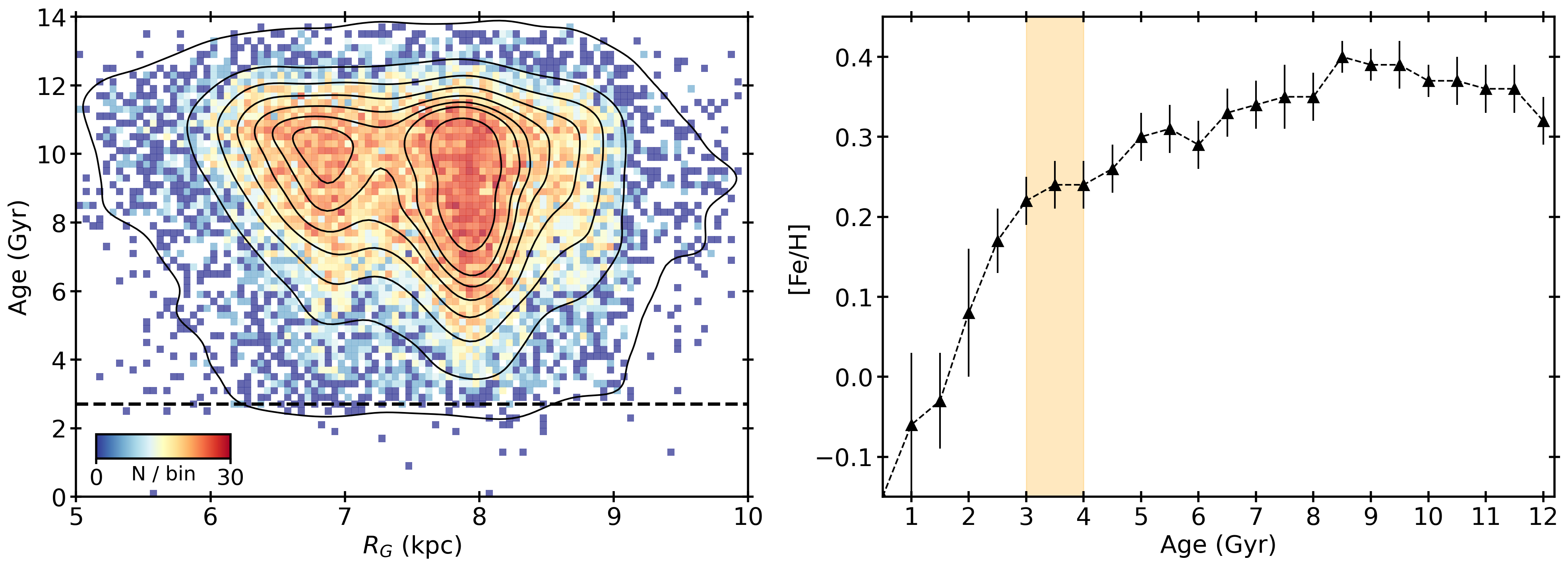}
    \caption{Left panel: Stellar ages as a function of guiding radii for stars with \feh>0.1 dex. The colors represent number of stars per bin. A Kernel Density Estimate (KDE) has also been over-plotted. A dashed black line at 2.7 Gyr is also shown. Right panel: Upper envelope of the age-metallicity relation traced by our full sample. The triangles represent the median \feh\, of 20 most metal-rich stars in each age bin and the error-bars show the scatter in \feh.}
    \label{fig:upper_envelope}
\end{figure*}

\subsection{Timing the bar formation epoch with the youngest super-metal-rich stars}\label{timing}

\begin{figure*}[!ht]
    \centering
    \includegraphics[width=0.8\linewidth]{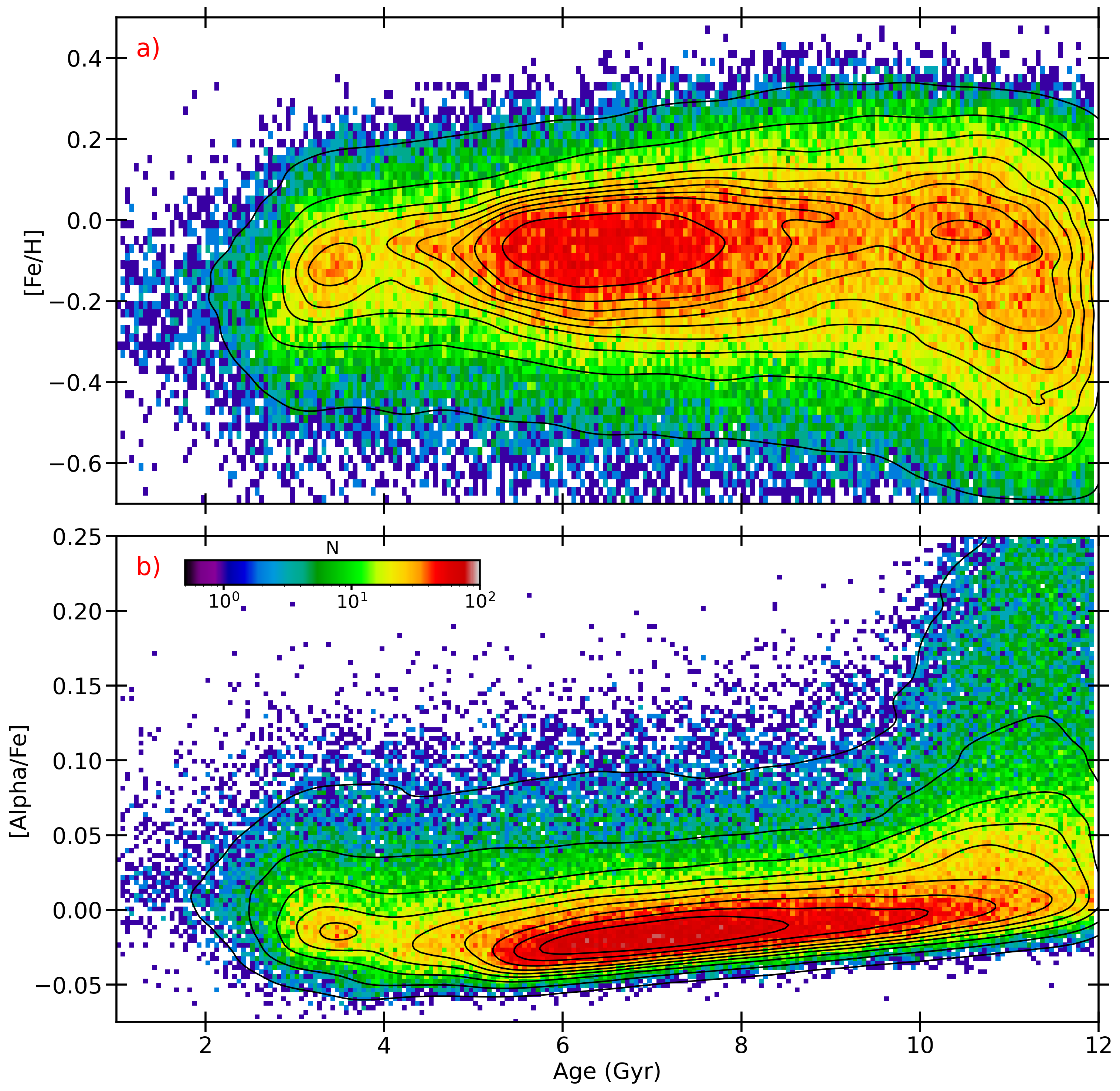}
    \caption{Top (a): 2D density distribution showing \feh\, as a function of stellar ages for the full sample. Bottom (b): 2D density distribution showing \alphafe\, as a function of stellar ages. A Kernel Density Estimate (KDE) have also been overplotted to highlight the density features in the AMR and AAR relationships. The colors represent number of stars per bin in log scale.}
    \label{fig:age_vs_mh_alpha}
\end{figure*}

In the left panel of Fig. \ref{fig:upper_envelope} we present the stellar ages as a function of \rg, for the 19\,367 metal-rich stars for which bimodality is clearly seen in Fig. \ref{fig:rguiding_bimodality}, i.e. for \feh>0.1 dex. The plot shows that the metal-rich stars have a wide range of ages with higher prevalence at older ages (11 to 6 Gyr). This hints to the fact that a large number of metal-rich stars were formed at early times in our Galaxy's history. The bimodal nature of \rg, with peaks at 6.9 and 7.9 kpc, is clearly seen for all ages and the spread in \rg\, is narrower for the younger stars.  Although some of these older stars could have already dispersed to the outer galaxy at earlier times through various processes such as merger events \citep[e.g.][]{Helmi2020ARA}, most of them are still in cold orbits. Interestingly, we see a near absence of metal-rich stars with ages younger than $\sim$2.7 Gyr old (see the dashed line in Fig. \ref{fig:upper_envelope}, also see Fig. \ref{fig:rg_vs_age_mh}). This cannot be attributed to selection effects on our sample such as color or magnitude limits, see Appendix \ref{ages}. Also, significant number younger stars with sub-solar \feh\, are present in our sample, see Figs. \ref{fig:age_vs_mh_alpha} and \ref{fig:rg_vs_age_mh}. Therefore, the dearth of younger SMR stars could be related to the mechanism that brought these stars formed in the inner Galaxy to the SNd.

In the right panel of Fig. \ref{fig:upper_envelope} we present the time evolution of the upper envelope of the metallicity for our whole sample. The plot reveals that our Galaxy quickly attained the highest levels of metallicity enrichment, levelling around \feh=0.4 dex, already at $\sim$10 Gyr ago. Then starting at 8.5 Gyr ago we observe a slow decreases, at $-$0.04 dex/Gyr, up to 4 Gyr. Then for a period of 1 Gyr, from 4 to 3 Gyr ago, we see a plateau at \feh=0.25 dex, highlighted by an orange band. After 3 Gyrs we see a steep decline in the \feh\, envelope at a rate of $-$0.15 dex/Gyr (see Appendix \ref{seismic ages} for a validation with an asteroseismic sample). This decline of the \feh\, envelope further confirms that the mechanism, that brought majority of these metal-rich stars to SNd, has ceased. We note here that \citet{Minchev2011} identified a radial migration mechanism, via bar and spiral structure interaction, that can migrate stars throughout the disc of MW-like galaxy in < 1 Gyr period. However, the migration induced by the bar is most effective at the epoch of bar growth and, therefore, stars in the inner Galaxy can be moved en masse to larger \rg\,\citep[e.g.][]{Halle2015}.

Considering MW bar growth and outward propagation of the resonances as the transport mechanism of these SMR stars to the SNd \citep{Minchev2010, Khoperskov2020_bar_escapees, Iles2023}, we can use the stellar ages of these youngest SMR stars to place a constraint on the epoch of intense bar growth following bar formation. It is worth noting that the MW bar, and inner Galaxy in general, when currently observed will contain stars of all ages from very old in higher number to very young, including those formed before, during and after the epoch of bar formation. However, only the youngest metal-rich stars, formed during the epoch of bar formation, can be confidently employed to constrain the age of bar. Based on this absence of younger SMR stars, we place the intense bar growth for the MW ending at 2.5 to 3 Gyrs ago.

\subsection{Enhanced star formation triggered by the bar formation}\label{starburst}

Considering that intense bar growth for MW ended at $\sim$3 Gyrs ago (see previous Section), we can expect an epoch of bar enhanced SF succeeding this event (e.g. \citealt{BabaKawata2020}). In this section, we explore SF in the Age-Metallicity and Age-Alpha Relationships (AMR and AAR) and look for possible imprints of such an event.

Fig. \ref{fig:age_vs_mh_alpha} shows, the AMR (panel a) and AAR (panel b) stellar density distributions for the whole sample. The AMR shows three distinct density features, at correspondingly three age regimes.

At the oldest regime, from 12 to $\sim$9-10 Gyrs, an oblique feature shows a steep rise in \feh\, with age with the broadest \feh\, distribution. This oblique feature depicts an epoch of intense SF that leads to a rapid chemical enrichment in the early MW from the lowest to the highest \feh\,, i.e. from $-$0.7 to 0.4 dex. In the accompanying AAR, at this regime, we see the high values of \alphafe\, corresponding to the SF burst in the early galaxy leading to the formation of the chemical thick disc and old bulge (e.g. \citealt{fuhrmann_1998, anders_2018, Miglio2021, Montalban2021NatAs, Queiroz2023, Xiang2022Natur}). In addition, this epoch is also attributed to the merger of MW with ~Gaia-Sausage-Enceladus \citep[GSE;][]{Belokurov2018, Helmi2018Natur}. 

Between 9 to 5 Gyrs we see a horizontal blob with most of the stars between $-$0.2<\feh<0.0. This epoch reflects a slow and steady SF and is attributed to the growth of the thin disc (e.g. \citealt{Chiappini1997, minchev_2013}). This regime shows a birth of a large number of stars with sub-solar \feh. A trend of decreasing \alphafe\, with age is seen, also a sign of slow but continued SF. Growth of the MW thin disc in an inside-out manner, due to steady gas infall is expected in this regime (see \citealt{Matteucci2021} and references therein).

Between 5 to 4 Gyrs we see a decline in SF, followed by another feature between 4 to 2.7 Gyr. This oblique feature shows an enhanced SF with a slight decrease in metallicity at younger ages. An increase in the \alphafe\, is also seen, showing a signature of SF burst. This result further confirms the enhanced SF during similar epoch previously reported \citep{Rocha-Pinto2000, Isern2019, Mor2019, Sysoliatina2021, Sahlholdt2022, Imig2023}. These mentioned previous works have not conclusively attributed any physical process to this SF phase. Considering that intense bar growth period ended at $\sim$3 Gyr ago, see Sect. \ref{timing}, we deduce that this star-burst is caused by the high activity of the bar evolution.

A likely scenario one could consider is that this high bar activity is simply a buckling of an old bar. But, there is still no consensus if the bar-buckling triggers SF in the disc (See \citealt{Fragkoudi2020} and for opposite view see \citealt{Debattista2006ApJ, Lokas2020}).

Interestingly, in their study of the outer bar region, \citet{Wylie2022} also find an abrupt decline of younger stars followed by a significant fraction of stars at 2-4 Gyr for the inner disc and ring  region (see their Fig. 3). Additionally, bar formation triggered star burst have been seen to typically last for a duration of $\sim$1 Gyr (e.g. see \citealt{BabaKawata2020}), this corroborates our assessment. Furthermore, bars have been observed to enhance SF in external galaxies. For example, see study by \cite{LinLin2020} on low-redshift galaxies using integral field spectroscopy (also see \citealt{Ellison2011MNRAS}). Hence, we conclude that MW bar formation occurs at $\sim$4 Gyr ago with end of strong bar activity at $\sim$2.7 Gyr ago. 

\section{Conclusion}\label{Section:conclusion}

We have explored a large and homogeneous sample of 169\,701 MSTO and SGB stars with 6D phase space information and high-quality stellar parameters coming from the \emph{Gaia}-DR3 RVS analysis of \citet{rvs_cnn_2023}. We supplemented the chemical abundances with stellar ages, distances and kinematics to study the epoch of MW bar formation. Thanks to \emph{Gaia} DR3 we obtain a mean distance uncertainty of 1\% which greatly contributes to the findings of this Letter.

The new data shows two new results: a) a clear bimodality in \rg\,at all ages (>3 Gyrs) of SMR stars, and b) a dearth of SMR stars younger than 3 Gyr. These results imply: 

\begin{itemize}
    \item Milky Way's bar had a strong activity phase lasting $\sim$1 Gyr ending at $\sim$3 Gyr ago. During the phase of strong bar activity, stars formed in the inner region (bulge/bar), are significantly redistributed across the outer disc, with the highest probability of migration around bar resonances. We verify this with a dearth of SMR stars younger than 3 Gyrs and an observed bimodality in the \rg\, around the bar resonances for the SMR stars, which are, with high confidence, formed in the inner galaxy and brought here during the strong activity phase of bar formation.
    \item We detect an enhancement in the global SF (around $-$0.3 < \feh < 0.0) at around 3 Gyr which, by that time had already declined in the local thin disc. Although this SF enhancement has been detected previously in the literature, the age coincidence with our estimate of the bar age suggest these events to be related.
    \item We suggest that, due to mixing and strong gas inflow due to bar there was an epoch of enhanced SF which is seen in Fig. \ref{fig:age_vs_mh_alpha}. This gas inflow during this phase lowers the upper floor of \feh\, vs age distribution while causing an increase in the \alphafe\, due to intense SF.
\end{itemize}

Future spectroscopic surveys such as 4MIDABLE-LR \cite{chiappini2019} will enable a large increase in the number of super metal rich stars with full 6D phase space and chemical information. In addition, the Japan Astrometry Satellite Mission for INfrared Exploration (JASMINE - \citealt{Kawata2023}) will be able to provide additional constraints on the age of the MW bar by precisely measuring the age of the Nuclear Stellar Disc. 

\begin{acknowledgements}
SN thanks the E-science \& IT team for COLAB service, computational clusters and research infrastructure at AIP. SN thanks Sergey Khoperskov, Friedrich Anders and Ivan Minchev for the helpful suggestions. GG acknowledges support by Deutsche Forschungs-gemeinschaft (DFG, German Research Foundation) – project-IDs: eBer-22-59652 (GU 2240/1-1). This project has also received additional funding from the European Research Council (ERC) under the European Union's Horizon  2020 research and innovation programme (Grant agreement No. 949173). APV acknowledges the DGAPA–PAPIIT grant IA103122. AM acknowledges support from the ERC Consolidator Grant funding scheme (project ASTEROCHRONOMETRY, G.A. n. 772293). This work has made use of data from the European Space Agency (ESA) mission \emph{Gaia}
(\url{https://www.cosmos.esa.int/gaia}), processed by the \emph{Gaia} Data Processing and Analysis Consortium (DPAC,
\url{https://www.cosmos.esa.int/web/gaia/dpac/consortium}). Funding for the DPAC has been provided by national institutions, in particular the institutions participating in the \emph{Gaia} Multilateral Agreement. This work made use of \texttt{overleaf} (\url{https://www.overleaf.com/}), and of the following \textsc{python} packages: \textsc{matplotlib} \citep{Hunter2007}, \textsc{numpy} \citep{Harris2020}, \textsc{pandas} \citep{mckinney-proc-scipy-2010}, \textsc{seaborn} \citep{Waskom2021}. This work also benefited from \textsc{topcat} \citep{Taylor2005}.
\end{acknowledgements}

\bibliographystyle{aa}
\bibliography{cite_b}

\begin{appendix}

\section{Details on input parameters for {\tt StarHorse} and kinematic calculations}\label{SH and galpy}

The distances and stellar ages are computed with the {\tt StarHorse} Bayesian isochrone-fitting method \citep{queiroz2018, Queiroz2023} as noted in Sect. \ref{data}. As inputs to {\tt StarHorse} we use the spectroscopic parameters from G23, galactic longitude (l) and latitude (b), photometric magnitudes \g, \bp\, $\&$\, \rp\, and parallaxes from \emph{Gaia} DR3 along with parallax corrections by \cite{Lindegren2021}. We also use the infra-red photometry (JHKs) from Two Micron All Sky Survey (2MASS; \citealt{2MASS}).

To calculate positions and velocities in the galactocentric rest-frame and to integrate the orbits of the stars, we use the 6D phase-space coordinates (sky positions, parallaxes, proper motions and radial velocities) from \emph{Gaia} DR3 \citep{gaiadr3_survey_properties} along with the {\tt StarHorse} distances. The integration of orbits was done with {\tt Galpy} \citep{galpy2015}, a python package for Galactic dynamics calculations. We use {\tt Astropy} \citep{astropy2022} for coordinate and velocity transformations, assuming the Sun is located at radius ~$\mathrm{R_{0}}$\,=\,8.2\,kpc and the circular velocity of Local Standard of Rest (LSR) as  ~$\mathrm{V_{0}}$\,=\,233.1\,\kms\,\citep{galpy2015, McMillan2017}. The peculiar velocity of the Sun with respect to the LSR is $\mathrm{(U, V, W)_\odot} = (11.1,\,12.24,\,7.25)$\,\kms \citep{Schonrich2010}. To run {\tt Galpy} we adopt the MW potential of \cite{McMillan2017} and perform orbit integrations for a 3 Gyr period and save each orbit's trajectory every 2 Myr.

The guiding radius we compute, given by \( R_g = L_Z / V_0\), is independent of the axisymmetric potential. Here, \(L_Z\) denotes the star's instantaneous angular momentum, defined as \(L_Z = R \cdot V_{\phi}\), where R is its galactocentric distance, and \(V_{\phi}\) denotes its azimuthal velocity in the Galactic plane.

\section{Kinematics of the metal-rich stars}\label{kine}

\begin{figure}[!ht]
    \centering
    \includegraphics[width=0.99\linewidth]{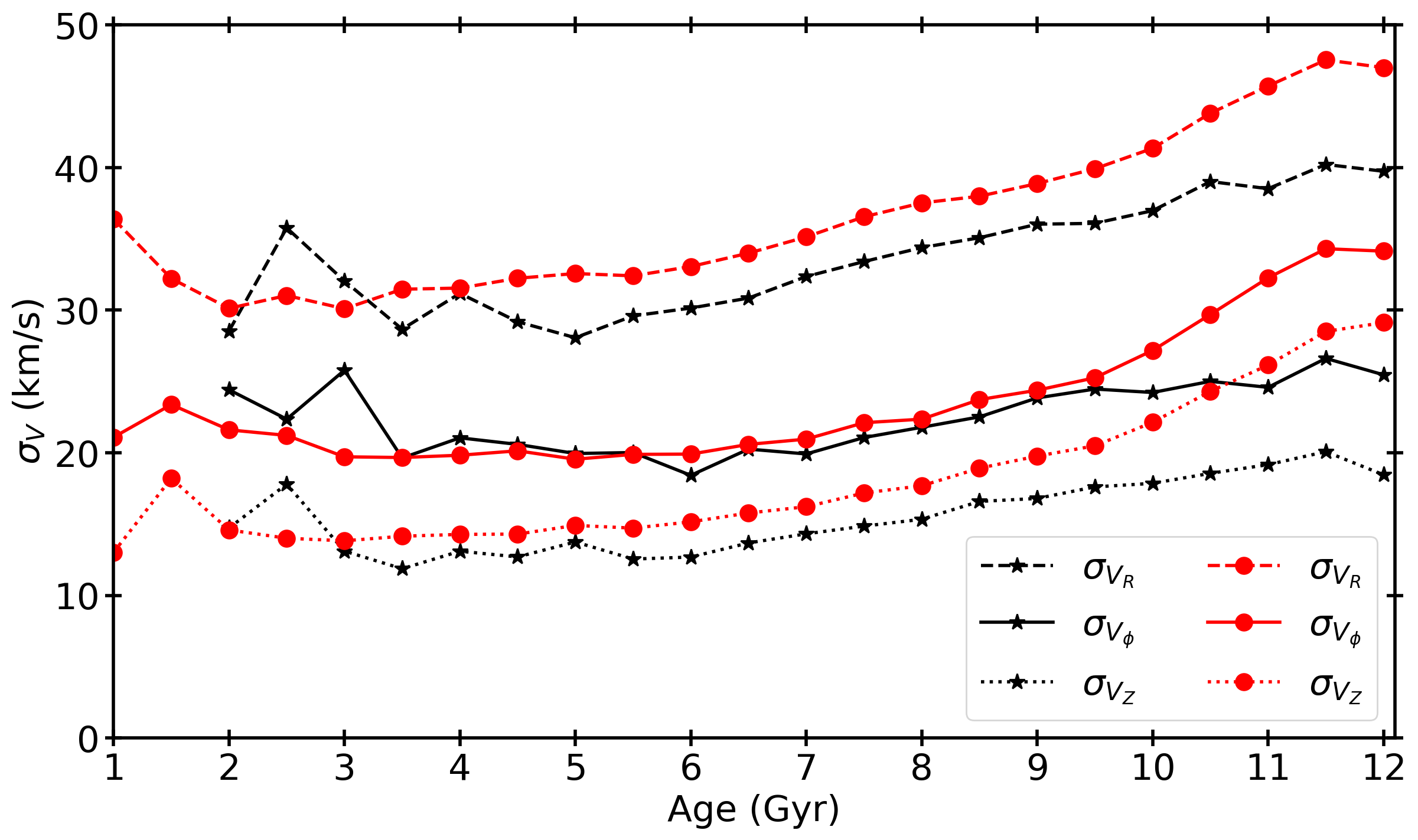}
    \caption{Velocity dispersions, radial ($\sigma_{V_R}$, dashed), azimuthal ($\sigma_{V_{\phi}}$, solid) and vertical ($\sigma_{V_{Z}}$, dotted), as a function of stellar ages. Red circles represent the full sample while black stars represent the metal-rich (\feh>0.1) stars.}
    \label{fig:vel disp}
\end{figure}

Here we compare the velocity dispersion of our entire sample (169\,727 stars) with that of stars with \feh>0.1. We highlight two main points in Fig. \ref{fig:vel disp}, namely: a) while $\sigma_{V_{\phi}}$ of the two samples are similar to each other in the 4-9 Gyr range, metal-rich stars have systematically lower $\sigma_{V_R}$ and $\sigma_{V_{Z}}$ (therefore systematically on cooler orbits than the bulk of the disc stars); b) at around 3 Gyr metal rich stars show a sudden increase in velocity dispersion. 

We also notice that for ages above circa 9-10 Gyr the sample traced by the red curves in Fig.\ref{fig:vel disp} shows a strong increase in velocity dispersion mainly due to the dominant contribution of thick disc stars. 

\section{Checking selection effects.}\label{ages}

\begin{figure}[!ht]
    \centering
    \includegraphics[width=0.9\linewidth]{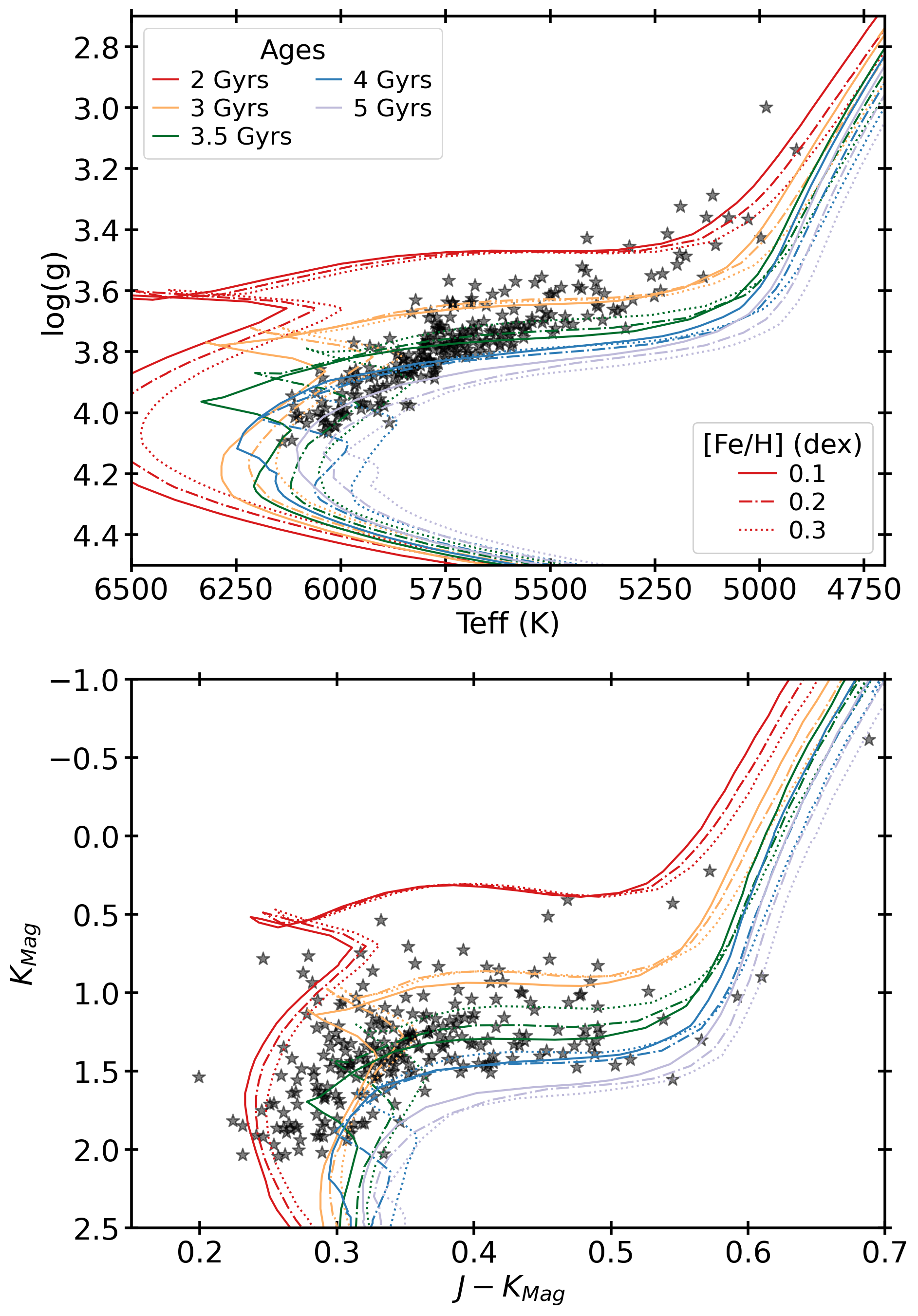}
    \caption{Kiel diagram and the CMD for the 256 metal-rich stars with 0.15 < \feh < 0.3 and 2 < Age (Gyr) < 4. For illustration purpose, PARSEC stellar isochrones, for ages 2, 3, 3.5, 4 \& 5 Gyrs and \feh = 0.1, 0.2 and 0.3 dex, are included. The CMD uses $JHKs$ absolute magnitudes from 2MASS which were used as inputs for {\tt StarHorse} to estimate ages. These stars are most probably formed during the epoch of bar formation.}
    \label{fig:evo_tracks}
\end{figure}

In order to clarify for the effects of any color/magnitude selection on the stellar ages we show, in Fig. \ref{fig:evo_tracks}, Kiel diagram (\teff\, vs \logg) and Color Magnitude Diagram (CMD) for a sample of 256 metal-rich stars with 0.15 < \feh < 0.3. PARSEC stellar isochrones \cite{PARSEC2012}, which are also used by {\tt StarHorse}, have been overplotted. The plots show a very well-behaved distribution of the metal-rich stars with no hard cuts on the color that hints at absence of young metal-rich stars. 

Given that our sample, obtained from \citet{rvs_cnn_2023}, is limited to \teff < 6300 K, and consists mostly of F, G, K stellar types, one could assume that we are biased against younger stars (more importantly younger SMR stars). Recently, \citet{Sahlholdt2022}, using a comparable data set of $\sim$ 180\,000 MSTO and SGB stars from the GALAH DR3 \citep{Buder2021}, study the age-metallicity distribution. Their sample covers a wider 8000 < \teff\,< 4000 range including hotter stars not present in our RVS-CNN sample (see their Fig. 1). Their AMR (see their Fig. 2) also shows that an absence of SMR stars younger than 2-3 Gyrs is real and not an effect of selection bias. This further supports our result on a dearth of SMR stars younger than 3 Gyrs ago as discussed in Sect. \ref{timing}. Also see Appendix \ref{seismic ages}. 

\section{Validation with external catalog.}\label{seismic ages}

\begin{figure}[!ht]
    \centering
    \includegraphics[width=0.9\linewidth]{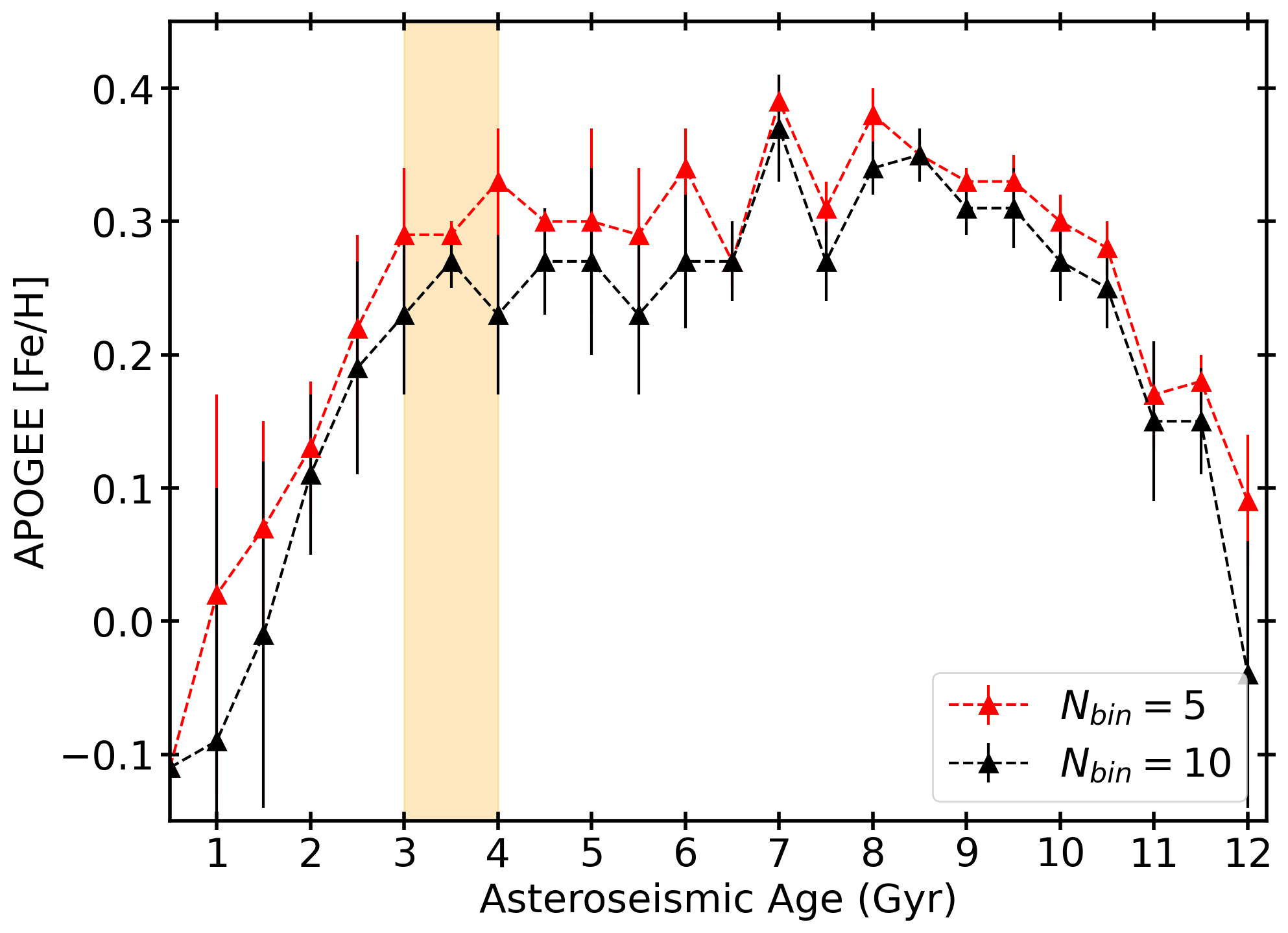}
    \caption{Upper envelope of the age-metallicity relation traced for the stars with asteroseismic ages from \citet{Miglio2021} and \feh\, from the APOGEE DR17 catalog \citep{Abdurrouf2022}. The triangles represent the median \feh\, of 5 (red) / 10 (black) most-metal-rich stars in each age bin and the error-bars show the scatter in \feh. Smaller number of stars per bin, for median and scatter, are used due to smaller number of stars in the \citet{Miglio2021} catalog.}
    \label{fig:seismic_feh_envelop}
\end{figure}

\begin{figure}[!ht]
    \centering
    \includegraphics[width=0.8\linewidth]{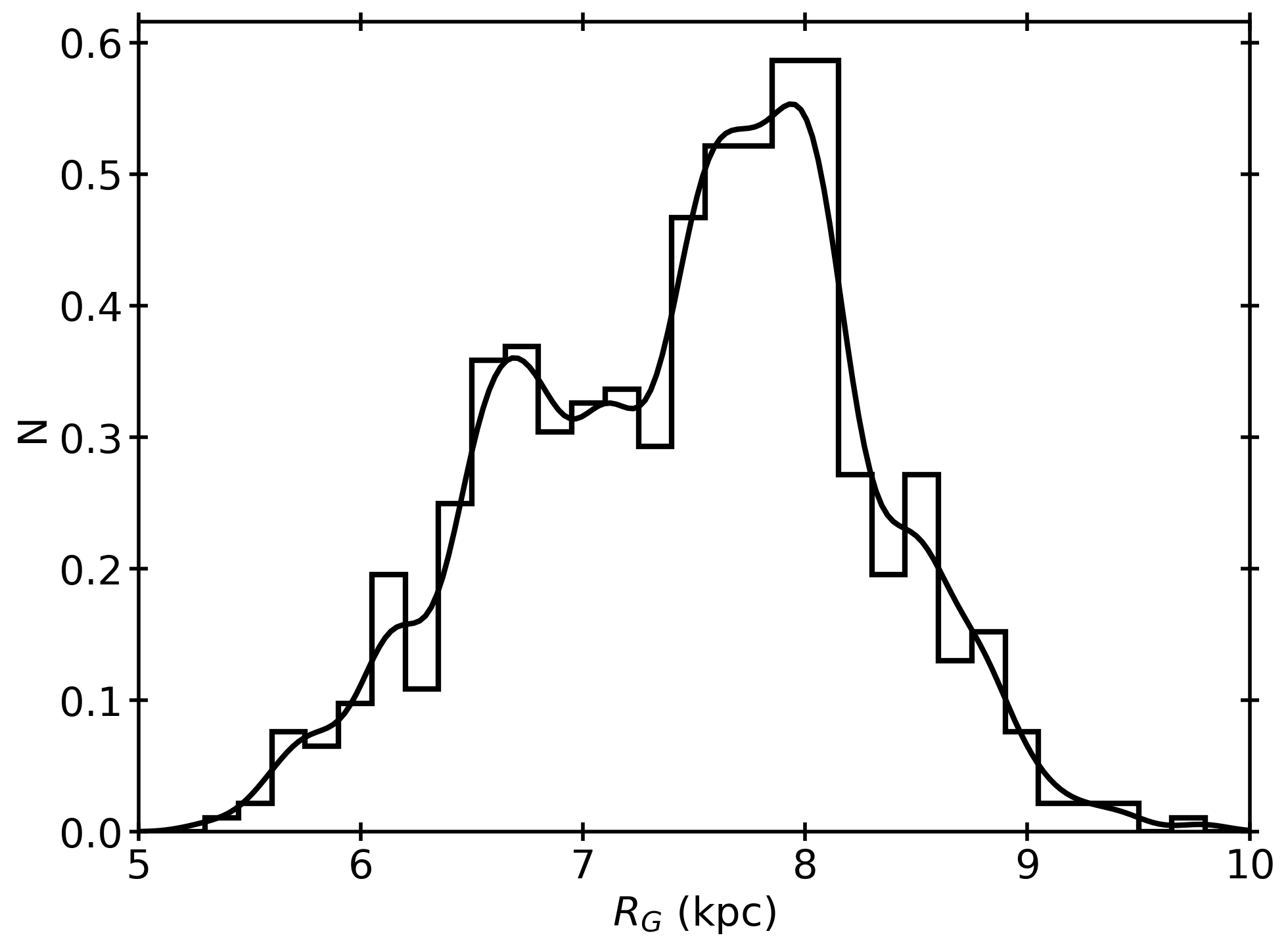}
    \caption{Distribution of the guiding radius (\rg) for the sample of metal-rich (\feh>0.1) red-giant stars with asteroseismic ages from the \citet{Miglio2021} catalog.}
    \label{fig:seismic_bimodality}
\end{figure}

Here, we perform the validation of our finding, of the steep decline of upper \feh\, envelope starting $\sim$3 Gyrs ago as discussed in Sect. \ref{timing}, using an external catalog. We use the Red-giant sample in the Kepler field from the \citet{Miglio2021} paper, reanalyzed with APOGEE DR17 parameters. After recommended quality cuts and selecting stars with age < 12 Gyr and \feh>-0.7, we obtained a sample of 2\,614 stars with mean age uncertainty of 25\%. Figure \ref{fig:seismic_feh_envelop} shows the time evolution of the upper envelope of the \feh\, for the asteroseismic sample. Similar to Fig. \ref{fig:upper_envelope}, the plot shows that our Galaxy attained high levels of \feh\, enrichment quite early with a slow decline until $\sim$ 3 Gyrs ago and then a steep decline after that. This confirms our finding of the dearth of SMR stars younger than 3 Gyrs old in the solar vicinity. In Fig \ref{fig:seismic_bimodality} we show that the metal rich stars (\feh>0.1) also show a bimodality in the guiding radius, similar to the finding with with the RVS-CNN sample as discussed in Sect. \ref{bimodality}.

\section{Guiding Radii vs Ages}\label{rg ages}

\begin{figure*}[!ht]
    \centering
    \includegraphics[width=0.8\linewidth]{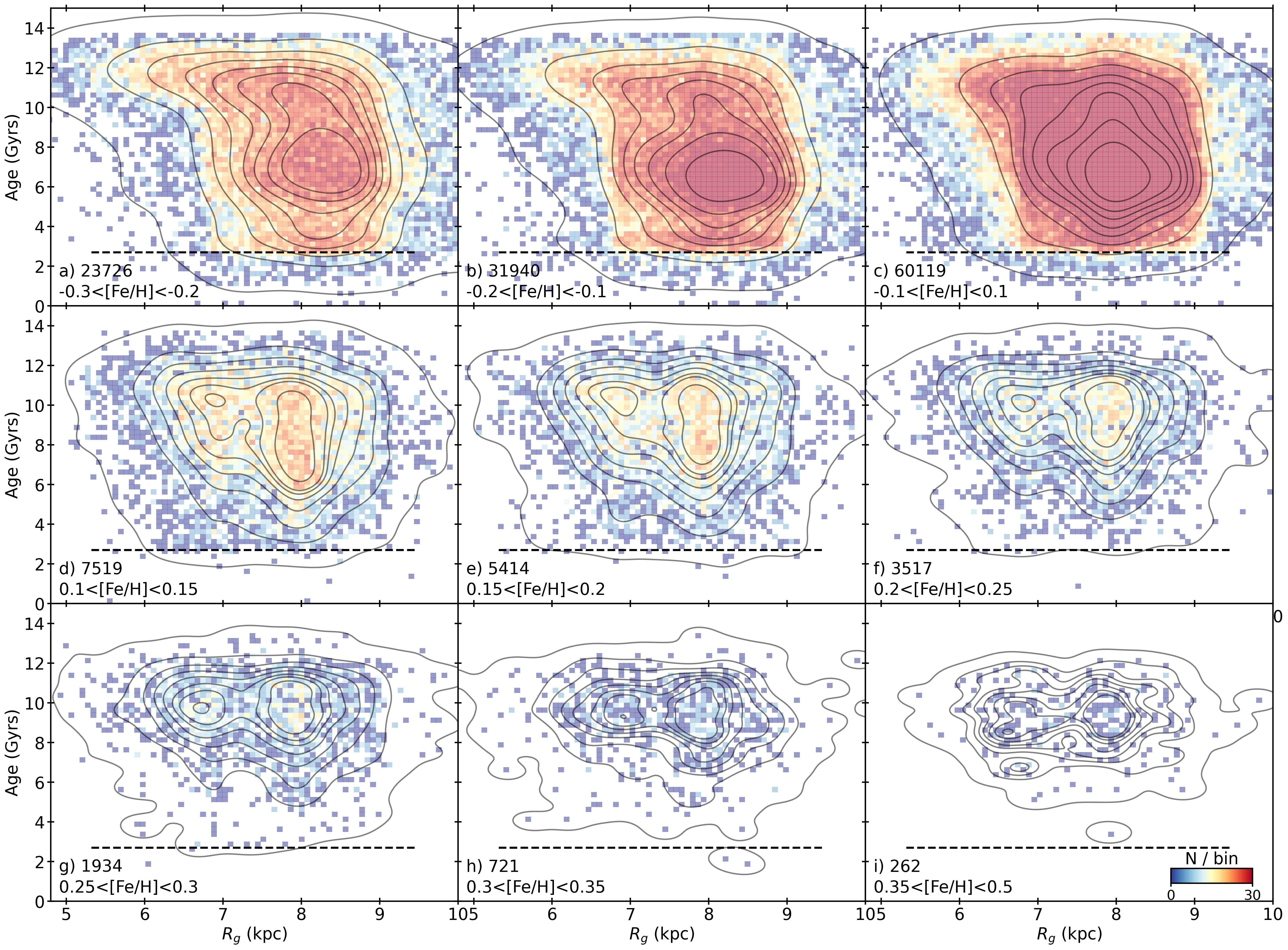}
    \caption{Stellar ages as a function of guiding radii, similar to left plot of Fig. \ref{fig:upper_envelope}, in \feh\, bins similar to Fig. \ref{fig:rguiding_bimodality}. A dashed black line at 2.7 Gyr is also shown. The colors represent number of stars per bin. A Kernel Density Estimate (KDE) has also been overplotted. As the \feh\, increases we find the bimodality in guiding radius appear more clearly. At solar and sub-solar metallicities we find a lot of young stars while for \feh>0.1 we see a near absence of stars younger than 2.7 Gyr old.}
    \label{fig:rg_vs_age_mh}
\end{figure*}

In Fig. \ref{fig:rg_vs_age_mh}, we present the stellar ages as a function of guiding radii, similar to left plot of Fig. \ref{fig:upper_envelope}, in \feh\, bins similar to Fig. \ref{fig:rguiding_bimodality}. The \feh\, values for each bins increases as we move from panel a) to panel i). The plot shows large number of stars, with a significant presence of stars younger than $\sim$3 Gyrs, for the sub-solar to solar metallicity bins while the \rg\, bimodality is not clear. As we move to \feh>0.1, the \rg\, bimodality becomes stronger with increasing \feh. The figure shows an absence metal rich (\feh>0.1) stars younger than 2.7 Gyrs, marked by the dashed line. We also see that the most metal-rich stars (\feh>0.3) are mostly old (11-6 Gyrs).

\section{Metal-rich stars as traces of bar activity} \label{bar tracers}

In Sect. \ref{bimodality} we discussed the \feh\, dependence of \rg\, bimodality revealed by the metal-rich stars. Here we test further to show how the metal-rich stars can trace bar activity. 

Substructures and ridges found in the SNd have been well studied and attributed to the bar (for detailed discussions see \citet{Kawata2018, Fragkoudi2019MNRAS, Chen2022}). Similarly, the Hercules Moving Group, which shows higher metallicity compared to the disc, has been proposed to consist of stars undergoing radial migration \citep{Liang_2023} and corotation of the bar has been suggested as the physical mechanism for its formation \citep[e.g.][]{Angeles2017,Monari2019,Khoperskov2022,Liang_2023}. 

\begin{figure*}[!ht]
    \centering
    \includegraphics[width=0.75\linewidth]{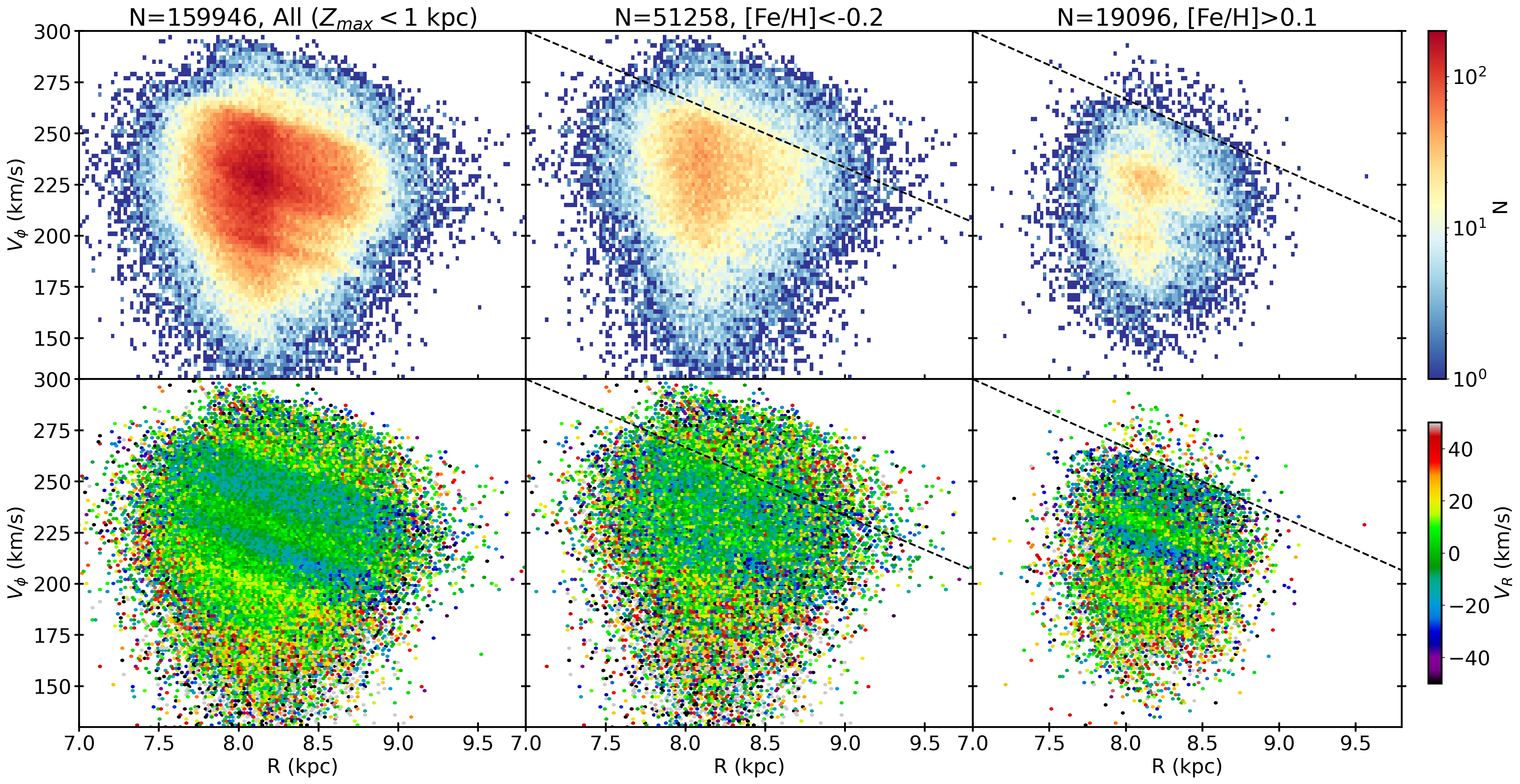}
    \caption{Top row: 2D density distribution showing the azimuthal velocity ($V_{\phi}$) as a function of galactocentric radius (R) for the disc stars (i.e. $Z_{max}<1$ kpc) to reveal the ridge-like features. Left panel shows all stars, mid panel shows metal-poor stars (\feh<$-$0.2) and right panel the metal-rich stars (\feh>0.1). Bottom row: Same plots as above shown to reveal the undulations when colored by the $V_R$. Both ridges and undulations are clearly seen for the full sample and the metal-rich stars while these features are not so clear for the metal-poor stars. In middle and right panels a dotted black line is drawn to represent the top sharp ridge seen in the top-left density plot.}
    \label{fig:R vs vphi}
\end{figure*}

In Fig. \ref{fig:R vs vphi}, we show our stars in the R vs $V_{\phi}$ plane to reveal the substructures in the phase-space. The density distribution, in top panels, clearly shows ridges in the phase space for the whole sample (thin disc stars with $Z_{max}$<1 kpc). These ridges are not visible for the metal-poor stars (mid panel) while they are clearly seen for the metal-rich stars. Furthermore, in the bottom panels we show how the undulations in $V_R$ are traced by our sample, primarily due to precise distances thanks to {\tt Gaia}. These undulations are very clear for the full sample and for the metal-rich set, however the signature is noisy for the metal-poor sample.

In Fig. \ref{fig:R vs vphi}, in middle and right panels, a black dashed line is drawn to show the location of the top sharp ridge seen in the full sample (top-left). We observe an interesting feature showing a near absence of metal-rich stars beyond this line while metal-poor stars are present. \citet{Khoperskov2022} using high-resolution N-body simulation, attribute this sharp decrease in mean \feh\, to the effect of OLR of the bar which limits the migration of the metal-rich stars, coming from inner Galaxy, beyond OLR.

These tests show that the metal-rich stars are indeed excellent tracers for bar activity.

\end{appendix}

\end{document}